%% file: arXiv2.tex
\documentclass[twocolumn,aps,prb,showpacs,preprintnumbers,amsmath,aimssymb]{revtex4-1}
\usepackage{graphicx}
\usepackage{bm}
\usepackage{color}
\usepackage{url}

\newcommand{\lf}{\left}
\newcommand{\rg}{\right}
\newcommand{\ra}{\rangle}
\newcommand{\la}{\langle}

\newcommand{\bea}{\begin{eqnarray}}
\newcommand{\eea}{\end{eqnarray}}
\renewcommand{\o}{\omega}

\begin{document}
\title{Interplay between destructive quantum interference and symmetry-breaking phenomena 
in graphene quantum junctions}

\author{A.~Valli,$^{1}$ A.~Amaricci,$^{1}$ V.~Brosco,$^{1,2}$ and M.~Capone$^{1}$}

\affiliation{$^{1}$Scuola Internazionale Superiore di Studi Avanzati (SISSA), 
and CNR-IOM Democritos, Istituto Officina dei Materiali, Consiglio Nazionale delle Ricerche, 
Via Bonomea 265, 34136 Trieste, Italy}
\affiliation{$^{2}$ISC-CNR and Department of Physics, Sapienza University of Rome, P.le A. Moro 5, 00185 Rome, Italy}

\pacs{}


\date{\today}
\begin{abstract}
We study the role of electronic spin and valley symmetry 
in the quantum interference (QI) patterns of the transmission function 
in graphene quantum junctions. 
In particular, we link it to the position of the destructive QI anti-resonances. 
When the spin or valley symmetry is preserved, 
electrons with opposite spin or valley display the same interference pattern. 
On the other hand, when a symmetry is lifted the anti-resonances are split, 
with a consequent dramatic differentiation of the transport properties 
in the respective channel. 
We demonstrate rigorously this link in terms of the analytical structure 
of the electronic Green function which follows from the symmetries 
of the microscopic model and we confirm the result 
with numerical calculations for graphene nanoflakes. 
We argue that this is a generic and robust feature 
that can be exploited in different ways for the realization of nanoelectronic QI devices, 
generalizing the recent proposal of a QI-assisted spin-filtering effect 
[A.~Valli et al. Nano Lett. ${\bf 18}$, 2158 (2018)].
\end{abstract}

\maketitle


\section{Introduction}
Quantum-interference (QI) effects in the electron transport in nanostructures
are a direct evidence of the particle-wave duality of electrons, 
which is deeply rooted in the fundamentals of quantum mechanics. 
From a theoretical point of view, 
it is well established that ballistic 
electron transport 
in molecular junctions characterized by multiple transmission paths 
displays clear signatures of QI. 
The prototype of completely destructive QI is the \emph{meta}-benzene 
molecular junction.\cite{solomonJCP129,cardamoneNL6,staffordNtech18} 
In the \emph{classical} interpretation, QI emerges when 
electrons propagating through two different spatial paths 
along the short- and the long-arms of the ring 
acquire a phase difference $\Delta\phi\!=\!\pi$,~\cite{cardamoneNL6,staffordNtech18}
yielding a complete cancellation of the transmitted wave amplitude. 
Interestingly, this view was recently challenged~\cite{nozakiJPC121} 
in favor of a different interpretation, where the antiresonance is a consequence 
of interference in energy space between different molecular orbitals. 
Independently of its origin, the presence of a QI antiresonance close to the Fermi level 
drastically influences the transport properties of quantum junctions
and results in huge ON/OFF ratios, which can be exploited for the realization of transistors~\cite{cardamoneNL6,staffordNtech18} 
or spin filters,~\cite{valliNL18,kaganJMMM440,protsenkoPRB99,liPRB99,li1906:01429} 
nanocircuitry,~\cite{calogeroJACS2019} 
and to enhance the thermoelectric performance~\cite{nooriNanoscale9} 
of nanoelectronic devices with organic functional units. 

Recently, experimental evidence of destructive QI was clearly observed in 
molecular junctions involving benzene,~\cite{yangCCL29} terphenyl,~\cite{liNatMat18} 
anthanthrene~\cite{gengJACS137}, antraquinone,~\cite{guedonNatNano7}, 
fullerenes and porphyrins,~\cite{gerlichNatComm2} 
as well as several other molecules with an organic backbone.~\cite{gantenbeinSR7} 
Sharp resonances in the differential conductance, 
the fingerprint of destructive QI, has been clearly detected 
even at room temperature.~\cite{batraFD174,familiChem5}
In some cases, the agreement between experiments and 
density functional theory calculations,~\cite{gengJACS137,markussenNL10,gantenbeinSR7} 
as well as with predictions made by graphical rules,~\cite{markussenNL10} is remarkable, 
thus establishing a scenario in which QI antiresonances 
can be regarded as robust features of quantum junctions, 
thus paving the way towards the realization of atomic-scale engineered 
quantum coherent devices. 

Poly-phenyl molecular systems, or, more generally, 
alternant hydrocarbons with delocalized $\pi$ orbitals 
represent the natural platform for QI effects. 
Remarkably, graphene nanostructures also fall into this category. 
Indeed, recent experiments reported QI patterns in 
graphene nanoconstrictions,~\cite{gehringNL16} or bridges,~\cite{nozakiJPCL6} 
and break junctions,~\cite{canevaNatNanotech13,batraFD174,familiChem5}  
What is more important, quantum junctions with graphene functional blocks 
benefit from all the extraordinary properties of graphene. 
Their chiral nature enables the manipulation of spin~\cite{slotaNat557} 
and valley~\cite{rycerzNP3,freitagNatNano5,suiNatPhys11,fujitaAPL97,grujicPRL113} 
degrees of freedom, while appropriate engineering of the substrate and gating 
offer the possibility to realize superlattices~\cite{ponomarenkoNat497} 
and to tune the properties of the junction. 
Furthermore, the presence of edges and reduced dimensionality offer the possibility 
to enhance correlation effects and to induce magnetic order, absent in pristine 
graphene,~\cite{chenSR3,magdaNat514,tucekNatComm8,slotaNat557} 
thus paving the way to a wide range of applications. 
Very recently, for instance, edge magnetism was stabilized 
in graphene nanoribbons functionalized with stable magnetic radical groups, 
demonstrating spin coherence times in the range of microseconds 
at room temperature.~\cite{slotaNat557}

The present work is related to all these aspects. 
By means of numerical calculations and a detailed symmetry analysis, 
we show that QI effects can be used to control spin and valley polarization 
of ballistic transport in graphene quantum junctions 
up to room temperature in the absence of external magnetic fields. 
In particular, we show that both spin filtering and valley filtering 
can be achieved in the same device by simply tuning the coupling 
with a substrate to switch the nature of the site-site correlations 
in the functional element between ionic and antiferromagnetic.

The paper is organized as follows. In Sec.~\ref{sec:GQJs} we discuss the 
model and the methods used to tackle the problem of correlated transport 
and QI effects in graphene nanostructures.
In Sec.~\ref{sec:SBS} we discuss the interplay between destructive QI 
and different kinds of symmetry-breaking, and we provide a unified description of the  phenomenon. 
In Sec.~\ref{sec:originQI} we explore the occurrence of the QI anti-resonances 
from a Green's function perspective, which allows us to pinpoint 
their origin. 
Finally, Sec.~\ref{sec:outlook} contains our conclusions and an outlook.

\section{Graphene quantum junctions}
\label{sec:GQJs}

\begin{figure}[t!]
\includegraphics[width=0.45\textwidth, angle=0]{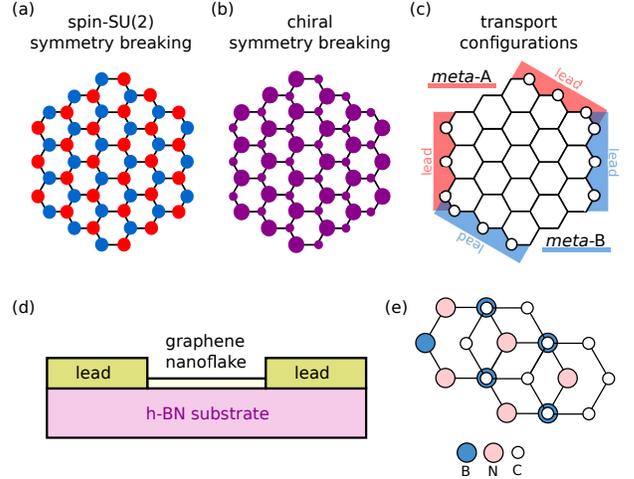}
\caption{(Color online) Schematic representations of 
(a) AF order in spin-SU(2) symmetry-broken case, where red (blue) circles 
represent spin-$\uparrow$ (spin-$\downarrow$) ordered magnetic moments, and 
(b) charge order in the chiral symmetry-broken case, 
where the size of the circle is proportional to the local charge density. 
(c) Two possible realizations of \emph{meta} contact configurations, 
i.e., at edge atoms from either sublattice A or sublattice B. 
(d) Quantum junction with a functional block consisting of a 
hexagonal graphene nanoflake with zigzag edges,
which is deposited on a h-BN substrate and is connected to two metallic leads. 
(e) Equilibrium stacking of the graphene/h-BN bilayer. }
\label{fig:junction}
\end{figure}

We consider the quantum junction schematically depicted in Fig.~\ref{fig:junction}. 
The Hamiltonian of the junction (${\cal H}$) can be decomposed in three terms,  
which describe the nanoflake (${\cal H}_F$), the leads (${\cal H}_L$), 
and the tunneling between the leads and the flake (${\cal H}_T$), respectively 
\begin{equation}
 {\cal H} = {\cal H}_F+{\cal H}_T+{\cal H}_L.
\end{equation}
The relevant features of the nanoflake are captured by the following 
low-energy effective Hubbard model 
for the delocalized $\pi$ electrons 
\begin{equation}
\label{eq:HHM}
\begin{split}
 {\cal H}_F = -t & \sum_{\langle ij \rangle\sigma} 
                p^{\dagger}_{i\sigma} p^{\phantom{\dagger}}_{j\sigma} 
                - \mu \sum_{i\sigma} n_{i\sigma} + U \sum_i n_{i\uparrow}n_{i\downarrow}\\
               + & \epsilon \sum_{\sigma} \Big( \sum_{i\in A} n_{i\sigma} 
                                              - \sum_{i\in B} n_{i\sigma} \Big).  
\end{split}
\end{equation}
where $p^{\dagger}_{i\sigma}$ ($p^{\phantom{\dagger}}_{i\sigma}$) 
create (annihilate) an electron at lattice site $i$ with spin $\sigma$, 
and $n_{i\sigma}\!=\!p^{\dagger}_{i\sigma} p^{\phantom{\dagger}}_{i\sigma}$ 
is the electron density operator.  
The parameter $t$ denotes the nearest-neighbor hopping amplitude on the honeycomb lattice, 
$\mu$ is the chemical potential, and the Hubbard $U$ describes the onsite Coulomb repulsion. 
Here, $\epsilon$ is an onsite energy that explicitly breaks the chiral symmetry 
between the A and B graphene sublattices, which can be induced, 
e.g., by the interaction between graphene 
and a suitable substrate, such as hexagonal boron-nitride (h-BN).

The metallic electrodes and the tunneling Hamiltonians, 
${\cal H}_L$ and ${\cal H}_T$, are instead given by
%
%
%
%
\begin{eqnarray}
\label{eq:HLT}
\begin{aligned}
 {\cal H}_L = & \sum_{\alpha k \sigma} 
                \epsilon^{\phantom{\dagger}}_{\alpha k\sigma} 
                c^{\dagger}_{\alpha k\sigma} c^{\phantom{\dagger}}_{\alpha k\sigma} , \\
 {\cal H}_T = & \sum_{\alpha ik\sigma}  \Big( 
                V^{\phantom *}_{\alpha ik\sigma}
                c^{\dagger}_{\alpha k\sigma} p_{i\sigma}^{\phantom{\dagger}} 
               + V^{*}_{\alpha ik\sigma} 
                p^{\dagger}_{i\sigma} c_{\alpha k\sigma}^{\phantom{\dagger}} 
                \Big) ,  
\end{aligned}
\end{eqnarray}
where the operators $c^{\dagger}_{\alpha k\sigma}$ ($c^{\phantom{\dagger}}_{\alpha k\sigma}$) 
create (annihilate) an electron with energy $\epsilon_{\alpha k\sigma}$ in lead $\alpha$, 
and $V_{\alpha ik\sigma}$ denotes the hopping amplitude 
between lattice site $i$ of the nanoflake and state $k$ of lead $\alpha$. 
We consider a hexagonal zig-zag edge graphene nanoflake with $N\!=\!54$ C atoms 
and a $C_3$ rotation symmetry around the center. 
As discussed in Ref.~\onlinecite{valliNL18}, destructive QI effects arise 
in contact configurations analogous to 
the \emph{meta} configuration of a benzene molecular junction. 
In the meta configuration for the nanoflake, 
the leads are connected at edge sites that belong to the same graphene sublattice. 
As depicted in Fig.~\ref{fig:junction}, there are two possibilities to realize such a configuration, i.e., when the edges belong to either the A or the B sublattice.

%
The Green's function of the nanoflake in the presence of the leads 
is obtained by solving the Dyson equation
\begin{equation}\label{eq:dyson}
 G^{-1}_{ij}(\omega) = G_{0,ij}^{-1}(\omega) 
                                 - \Sigma^{L}_{ij}(\omega) - \Sigma^{R}_{ij}(\omega) 
                                 - \Sigma_{ij}(\omega),
\end{equation} 
where $G_{0,ij}^{-1}(\omega)$ is the bare Green's function of the isolated nanoflake, 
which is renormalized by three self-energy contributions: 
$\Sigma^{L}_{ij}(\omega)$ and $\Sigma^{R}_{ij}(\omega)$, which describe 
the embedding of the nanoflake with the left (L) and right (R) lead, respectively,  
and $\Sigma_{ij}(\omega)$, which describe the electronic correlations 
stemming from the Hubbard interaction $U$ within the nanoflake.  

\noindent The leads contribution to the self-energy is given by
\begin{equation}
\label{eq:rethyb}
  \Sigma^{\alpha}_{ij\sigma}(\omega) = 
  \sum_{k} \frac{V^{\phantom *}_{\alpha ik\sigma}V^{*}_{\alpha jk\sigma}}
                {\omega\!+\!\imath\eta - \epsilon_{\alpha k\sigma}}, 
\end{equation} 
where $\eta\!>\!0$ regularizes the analytic continuation. 
The product $V^{\phantom *}_{\alpha ik}V^{*}_{\alpha jk}$  in Eq. (\ref{eq:rethyb}) describes 
virtual hopping processes in which an electron 
from site $i$ of the nanoflake is injected into state $k$ of lead $\alpha$ 
and (after a certain time) returns to site $j$ of the nanoflake. 
In the following, we restrict to \emph{local} hybridization processes, 
i.e., $\Sigma^{\alpha}_{ij}\propto\delta_{ij}$, 
without affecting the qualitative results presented below. 
Moreover, since the QI properties originate from the topology of the nanoflake, 
and are independent of the details of the coupling with the leads, 
it is reasonable to assume a wideband limit (WBL) approximation 
for the leads,~\cite{verzijlJCP138} 
in which $\Sigma^{\alpha}_{ii}\!=\!-\imath\Gamma$ 
and it is independent of energy for each contact site $i$. 

\noindent The effects of electronic correlations within the nanoflake 
are taken into account within the dynamical mean-field theory~\cite{georgesRMP68} (DMFT), 
in a real-space extension suitable to deal with systems 
where the translational symmetry is broken in one or more spatial 
dimensions.\cite{snoekNJP10,valliPRL104,jacobPRB82,valliPRB86,mazzaPRL117,schuelerEPJST226} 
This approach is also suitable to treat inhomogeneous systems 
in the presence of charge-~\cite{dasPRL107,valliPRB92} 
and spin-order,~\cite{snoekNJP10,valliPRB94,valliNL18,amaricciPRB98} 
which will be of importance in the following, 
as well as superconductivity.~\cite{amaricciPRA89}
Within real-space DMFT, the nanoflake 
is mapped onto a set of self-consistent auxiliary Anderson impurity problems, 
which are solved with a L\'{a}nczos exact diagonalization procedure,~\cite{caffarelPRL72,caponePRB76} 
yielding a \emph{local} yet site-dependent self-energy 
$\Sigma_{ij}(\omega)\!=\!\Sigma_{i}(\omega)\delta_{ij}$. 
Within this approximation, local quantum fluctuation are treated non-perturbatively, 
whereas non-local spatial correlations are retained at a static mean-field level. 

At the same time, DMFT also allows us to take into account 
finite temperature effects within the Green's function formalism.  
This includes the broadening of the Fermi-Dirac distribution function 
as well as non-trivial effects of the temperature evolution of the many-body states. 
In the following, we consider temperature effects at 
$T\!=\!0.005t$, which corresponds to $T\!\approx\!150$~K 
for a realistic value of $t\!=\!2.7$~eV of the nearest-neighbor hopping 
in graphene.~\cite{castronetoRMP81}

Starting from the Green's function of the nanoflake, 
under appropriate assumptions,~\cite{meirPRL68,nessPRB82,droghettiPRB95} 
the transmission of the junction can be estimated 
with the following Landauer-B\"{u}ttiker expression~\cite{landauerJRD1,buettikerPRL57}
\begin{equation}\label{eq:landauer}
 T(\omega) = \textrm{Tr}\Big[ \Gamma^{L}G^{a}\Gamma^{R}G^{r}\Big],
\end{equation}
where $G^{r(a)}$ is the retarded (advanced) Green's function 
obtained from Eq.~(\ref{eq:dyson}), while the matrix 
$\Gamma^{\alpha}\!=\!\imath\big[ \Sigma^{\alpha}-\Sigma^{\dagger \alpha} \big]$ 
encloses the spectral information of the leads. 
Within our description of the leads,  
the transmission in Eq.~(\ref{eq:landauer}) can be recast as
\begin{equation}\label{eq:trans_independent}
 T(\omega) = \sum_{\ell} \sum_{r} \Gamma^{L}_{\ell \ell} \ \Gamma^R_{rr} |G_{\ell r}^{r}(\omega)|^2, 
\end{equation} 
which represents a sum over \emph{independent} transmission channels,
with $\ell$ and $r$ labeling the lattice sites of the nanoflake 
connected to the L and R leads, respectively. 

As is evident from Eq.~(\ref{eq:trans_independent}), 
the energy dependence of the transmission 
is controlled entirely by the Green's function, thus establishing 
a direct relation between the transport properties of the junction 
and the electronic properties of the graphene nanoflake. 
It can be explicitly shown that corrections to the transmission function 
beyond the WBL do not change qualitatively 
the results presented in the following.~\footnote{See also the extended discussion 
in the Supporting Information of Ref.~\onlinecite{valliNL18}, 
at \protect\url{https://pubs.acs.org/doi/abs/10.1021/acs.nanolett.8b00453}} 

This is particularly relevant because it is possible to link the existence 
of destructive QI to the symmetries and the analytic properties of the Green's function 
(see Sec.~\ref{sec:originQI} and Appendix~\ref{app:a}). 
This suggests that the extraordinary filtering properties 
of the device in the spin- and valley- channels are robust features, 
which depend neither on the details of the nanoflake 
and of the lead-flake hybridization, 
nor on the approximation employed in the calculations.~\cite{valliNL18}

\section{Results and Discussion}
\label{sec:SBS}

In the following we discuss the electronic and transport properties 
of the hexagonal graphene nanoflake quantum junction. 
In particular, we focus on the interplay between 
the destructive QI in the meta configuration and the symmetry breaking phenomena
involving the spin- and valley- degrees of freedom. 

\subsection{Effects of the symmetry breaking on a destructive QI antiresonance}
To realize the scenario in which we are interested, 
the minimal requirements for the transmission function $T_{\lambda}(\omega)$ are as follows: 
(i) $T_{\lambda}(\omega)$ displays a QI antiresonance at 
$\omega=\omega_{\lambda}^{\textrm QI}$ 
in a given channel, denoted by $\lambda$, which has two (or more) components; 
(ii) $T_{\lambda}(\omega)$ is the same for each component of $\lambda$, 
at least close to $\omega_{\lambda}^{\textrm QI}$, 
when the symmetry associated with $\lambda$ is not broken. 

\begin{figure}[t]
\includegraphics[width=0.47\textwidth, angle=0]{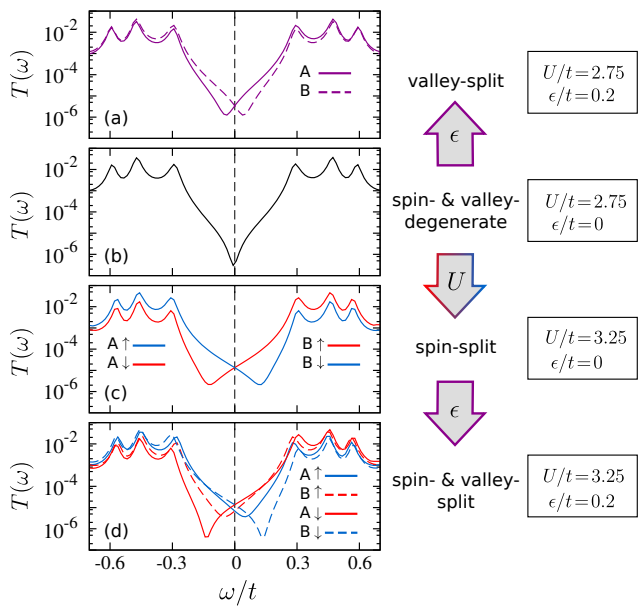}
\caption{(Color online) Evolution of $T_{\lambda}(\omega)$ in the spin- and valley- transmission channels. 
By breaking the spin-$SU(2)$ or the chiral symmetry (or both) the four-fold degeneracy ($g_{\lambda}\!=\!4$) 
of the QI antiresonance at $\omega^{\textrm QI}\!=\!0$ is lifted, 
and $g_{\lambda}\!=\!2$ or $g_{\lambda}\!=\!1$ QI antiresonances 
at $\omega_{\lambda}^{\textrm QI}\!\neq\!0$ 
appear in the corresponding channel. The curves of each case are shifted vertically for clarity. 
Parameters: $\Gamma/t=0.02$, $T/t\!=\!0.005$, while $U/t$ and $\epsilon/t$ as labeled.}
\label{fig:Te_evo}
\end{figure}

In the present case, the previous requirements are fulfilled, 
in any of the meta contact configurations of the junction, 
for both the spin and valley pseudo-spin 
(i.e., $\lambda\!=\!\{\sigma,\tau\}$, with $\sigma\!=\!\pm 1$ and $\tau\!=\!\pm 1$). 
When the $SU(4)$ symmetry associated with the combined degrees of freedom is intact, 
we observe a QI antiresonance with multiplicity $g_{\lambda}=4$, 
while breaking the spin- or the chiral- $SU(2)$ symmetry (or both) results 
in a lifting of the degeneracy of $\omega_{\lambda}^{\textrm QI}$ 
and a strong differentiation of the transport properties 
due to the destructive QI. 
We summarize our findings in Fig.~\ref{fig:Te_evo}, 
where we show how the transmission $T(\omega)$ changes 
when breaking the symmetries of the Hamiltonian. 
When neither the spin nor the valley degeneracy is lifted, $T(\omega)$ is the same 
in all channels and displays a four-fold antiresonance 
at $\omega^{\textrm QI}\!=\!0$, signature of destructive QI. 
This scenario is depicted in Fig.~\ref{fig:Te_evo}(b), 
while Fig.~\ref{fig:Te_evo}(a, c, and d) correspond 
to all the different symmetry-breaking scenarios,  
which we are going to discuss below in details.

\subsection{Spin-split scenario}
\begin{figure}
\includegraphics[width=0.47\textwidth, angle=0]{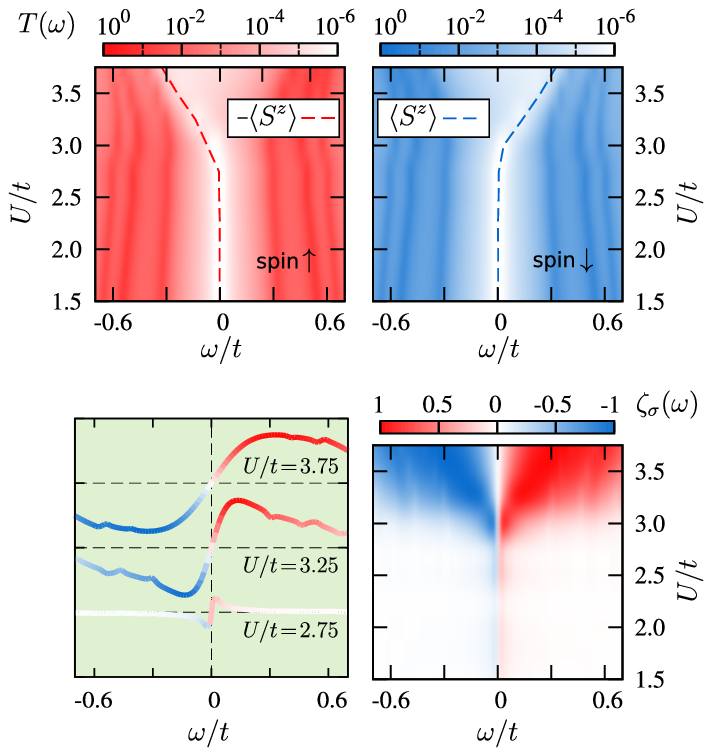}
\caption{(Color online) Interplay between destructive QI 
and spin-$SU(2)$ symmetry breaking. 
Map of the spin-resolved transmission $T_{\sigma}(\omega)$ (top panels). 
The dashed lines show the relation 
$\omega^{\textrm QI}_{\sigma}\!\approx\!\sigma \langle S^z \rangle$, 
while the gap can be estimated by the energy difference of the lowest-energy 
(HOMO and LUMO) transmission resonances. 
Map of the spin polarization $\zeta_{\sigma}(\omega)$ and corresponding cuts 
at different values of $U/t$ (lower panels).
Parameters: $\epsilon/t\!=\!0$, $\Gamma/t\!=\!0.02$, and $T/t\!=\!0.005$. 
The data shown are for valley B, but the results are analogous for valley A, 
when the proper symmetry relations are considered, as discussed in the text.}
\label{fig:Te_spin}
\end{figure}

The ground state of the nanoflake changes from paramagnetic (PM) 
to antiferromagnetic (AF) when the local repulsion 
overcomes a critical threshold ($U\!>\!U^{AF}$). 
The AF state breaks the spin-$SU(2)$ symmetry, 
with a N\'{e}el-like pattern of ordered moments 
$\langle S^z_{i} \rangle \!=\! \langle n_{i\uparrow} \rangle \!-\! \langle n_{i\downarrow} \rangle$,  
which have opposite polarization in the two graphene sublattices, 
and a finite staggered magnetization 
\begin{equation}\label{eq:Sz}
 \langle S^z \rangle \!=\! \frac{1}{N} \Big( \sum_{i \in A} \langle S^z_{i} \rangle - \sum_{i \in B} \langle S^z_{i} \rangle \Big).
 \end{equation} 
The local magnetic moments are spatially inhomogeneous, and they increase 
with the distance from the center of the nanoflake.~\cite{rossierPRL99,kabirPRB90,valliPRB94}
The magnetic pattern is stabilized by short-range antiferromagnetic correlations, 
which are stronger at the edges and weaker in the bulk.\cite{valliPRB94} 
In the spin-$SU(2)$ symmetry-broken state, the transmission for a given valley
is no longer  the same in the spin-$\uparrow$ and spin-$\downarrow$ channels. 
The spin-resolved transmission $T_{\sigma}(\omega)$ still exhibits destructive QI, 
but the antiresonances are separated in energy and located at 
$\omega^{\textrm QI}_{\sigma}\!\propto\!\sigma\langle S^z\rangle$, 
as shown in Fig.~\ref{fig:Te_evo}(c) and in Fig.~\ref{fig:Te_spin} explicitly. 

The selective suppression of the transmission in one of the spin channels, due to destructive QI, 
can be exploited to obtain a \emph{nearly perfect} QI-assisted spin-filtering device, 
as recently proposed in Ref.~\onlinecite{valliNL18}, 
which demonstrates the potential impact 
of the investigated phenomenon for technological applications. 
Note that, since the two sublattices have opposite magnetization, 
the transmission is still symmetric under the simultaneous inversion 
of the spin ($\sigma\!\rightarrow\!\bar{\sigma}$) and valley pseudospin ($\tau\!\rightarrow\!\bar{\tau}$) 
i.e., $T_{\tau\sigma}(\omega)\!=\!T_{\bar{\tau}\bar{\sigma}}(\omega)$, 
as specified in the legend of Fig,~\ref{fig:Te_evo}. 
This means also that the QI antiresonance is still two-fold degenerate 
$\omega^{\textrm QI}_{\tau\sigma}\!=\!\omega^{\textrm QI}_{\bar{\tau}\bar{\sigma}}$.

The transport properties of the junction in this scenario are shown 
in details in Fig.~\ref{fig:Te_spin}. 
The top panels show the heatmap of $T_{\sigma}(\omega)$ 
as a function of $\omega/t$ and $U/t$ 
separately for the spin-$\uparrow$ and spin-$\downarrow$ channels 
(for valley B, but the situation is analogous for valley A, as discussed above). 
From the point of view of the electronic structure, 
the information enclosed in the transmission function is equivalent 
to that of the electronic excitation spectrum. 
Indeed, one can follow the evolution of the electronic resonances 
(darker shades of color in the heatmap) 
and in particular, of the ones closest to the Fermi level, 
corresponding to the ighest occupied molecular orbital 
and the lowest unoccupied molecular orbital, which identify the spectral gap. 
At $U/t\!<\!U^{AF}/t\!\lesssim\!3$, the gap is \emph{reduced} 
with respect to its non-interacting value~\cite{valliPRB94}
as $\Delta\!\approx\!\langle Z \rangle \Delta_0$, 
where $\langle Z \rangle$ is the average over the nanoflake 
of local quasi-particle residue, extracted from the local DMFT self-energy as 
\begin{equation}
 Z_i = \Big( 1-\frac{\partial\Sigma_i(\omega)}{\partial\omega}
                             \Big|_{\omega\rightarrow 0} \ \Big)^{-1}.
\end{equation} 
Instead, when AF sets in, the gap is no longer controlled 
by the quantum confinement effect, but by the staggered magnetization, 
and it \emph{increases} with $U$. 
While the transmission is exponentially suppressed within the energy gap, 
destructive QI manifests itself in the form of a QI antiresonance 
(in the middle of the white area in the heatmap). 
In the PM state, the anti-resonance is \emph{pinned} 
at the Fermi level $\omega^{\textrm QI}\!=\!0$ 
due to the particle-hole symmetry of the spectrum.\cite{pedersenPRB90,valliNL18}
In the AF state instead $\omega^{\textrm QI}_{\sigma}$ is spin-dependent 
and shifts below or above the Fermi level, proportionally 
to the average staggered magnetization $\langle S^z \rangle$. 
The detailed analysis of the transmission as a function of $U/t$ 
shown in Fig.~\ref{fig:Te_spin} explains the change in the transmission 
Figs.~\ref{fig:Te_evo}(b,c) in between the symmetric and the spin-split scenarios.

\begin{figure}
\includegraphics[width=0.47\textwidth, angle=0]{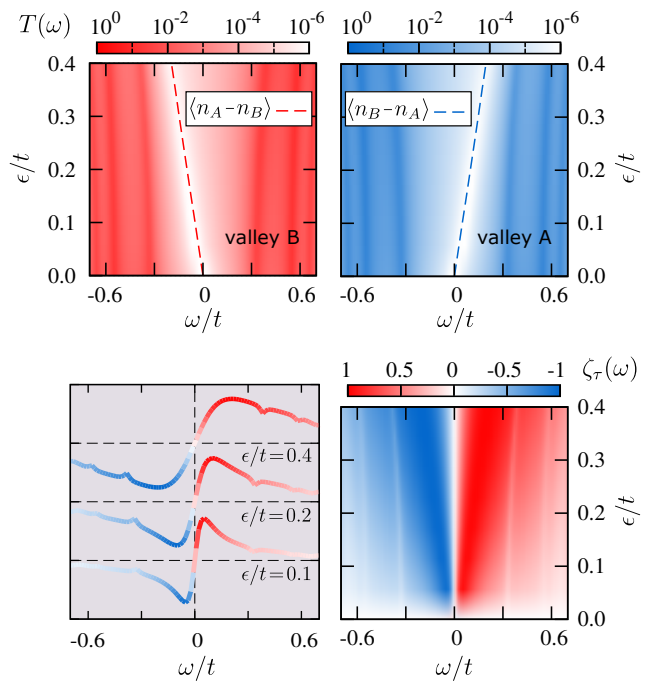}
\caption{(Color online) Interplay between destructive QI 
and chiral symmetry breaking. 
Map of the valley-resolved transmission $T_{\tau}(\omega)$ (top panels).  
The dashed lines show the relation 
$\omega^{\textrm QI}_{\tau}\!\approx\!\tau \langle n_A-n_B \rangle$, 
while the gap can be estimated by the energy difference of the lowest-energy 
(HOMO and LUMO) transmission resonances. 
Map of the pseudo-spin polarization $\zeta_{\tau}(\omega)$ and corresponding cuts 
at different values of $\epsilon/t$ (lower panels). 
Parameters: $U/t\!=\!1.5$, $\Gamma/t\!=\!0.02$, and $T/t\!=\!0.005$.
The data shown are for the spin-$\uparrow$ channel, but the result are identical 
for the spin-$\downarrow$ channel, due to the spin-$SU(2)$ symmetry, as discussed in the text.}
\label{fig:Te_valley}
\end{figure}

As a consequence of the different behavior of 
$T_{\uparrow}(\omega)$ and $T_{\downarrow}(\omega)$, 
the spin-polarization of the transmission 
\begin{equation}\label{eq:zetas}
 \zeta_{\sigma}(\omega) = \frac{T_{\uparrow}-T_{\downarrow}}{T_{\uparrow}+T_{\downarrow}}, 
\end{equation}
is not zero in a wide frequency range above and below the Fermi level. 
This is demonstrated in the bottom panels of Fig.~\ref{fig:Te_spin}, 
where we show the heatmap of $\zeta_{\sigma}(\omega)$ as a function of $\omega/t$ and $U/t$, 
as well as cuts of $\zeta_{\sigma}(\omega)$ for specific values of $U/t$. 
In all cases, the maxima (or minima) of the polarization located at frequencies 
$\omega_{\sigma}^{\textrm QI}$, 
where the transmission probability in one spin channel is strongly suppressed 
and the transport is dominated by the other channel, 
thus achieving \emph{nearly perfect} spin filtering.~\cite{valliNL18}

\subsection{Valley-split scenario}
Let us consider the case in which $U\!<\!U^{AF}$ 
and the ground state of the system is therefore PM, 
and let us introduce a field that breaks the chiral symmetry, 
associated with the chemical equivalence of the graphene sublattices. 
Here we assume that this term originates from the interaction between graphene 
and a h-BN substrate. 
Indeed, \emph{ab-initio} density functional theory calculations 
have shown that in graphene/h-BN bilayers, in the equilibrium stacking position, 
the atoms of one of the two graphene sublattices iare stacked on top of B atoms, 
while the atoms of the other sublattice are located in the hollow position 
of the underlying h-BN honeycomb lattice.\cite{giovannettiPRB76} 
Following Skomski \emph{et al.},~\cite{skomskiMH563} 
the asymmetric adsorption of the graphene sublattices can be 
encoded in a single-particle term $\epsilon$, 
which has the form given in Eq.~(\ref{eq:HHM}) in the Hamiltonian of the nanoflake.

The chiral symmetry-breaking field $\epsilon$ induces 
a charge-density wave (CDW)
and drives the C atoms \emph{locally} away from half-filling 
(while the electric charge is overall conserved). 
At $\epsilon\!\neq\!0$ the two valleys are no longer degenerate 
and each of the valley-resolved transmission functions $T_{\tau}(\omega)$, 
shown in Fig.~\ref{fig:Te_evo}(a),   
displays a destructive QI antiresonance at 
$\omega^{\textrm QI}_{\tau}\!\approx\!\tau\langle n_A\!-\!n_B\rangle$, 
where 
\begin{equation}
 \langle n_{A/B} \rangle = \frac{2}{N}\sum_{\sigma}\sum_{i\in A/B} 
                           \langle n_{i\sigma} \rangle
\end{equation} 
is the sublattice-resolved charge density. 
Note that, analogously to the previous case, $T_{\tau}(\omega)$ retains 
a two-fold degeneracy due to the spin-$SU(2)$ invariance, and therefore 
$\omega^{\textrm QI}_{\tau\sigma}\!=\!\omega^{\textrm QI}_{\tau\bar{\sigma}}$. 

The analysis of the transport properties is presented in Fig.~\ref{fig:Te_valley} 
and can be done in complete analogy with that of the spin-split scenario. 
For this reason, we mostly focus on the differences between the two cases.
The heatmap of $T_{\tau}(\omega)$ as a function of $\omega/t$ and $\epsilon/t$ 
shows that the spectral gap \emph{increases} for any $\epsilon\!>\!0$, 
and $\omega^{\textrm QI}_{\tau}$ shifts away from the Fermi level 
proportionally to the charge-density wave order parameter. 
In contrast to the previous case, there is no finite critical threshold 
for the onset of the charge-density wave. 
The valley-polarization of the transmission $\zeta_{\tau}(\omega)$, 
defined as
\begin{equation}\label{eq:zetav}
 \zeta_{\tau}(\omega) = \frac{T_A-T_B}{T_A+T_B} 
\end{equation}
is finite for any $\epsilon\!\neq\!0$. 
The maxima (or minima) of the polarization $\zeta_{\tau}(\omega)$ are always located at frequencies $\omega_{\tau}^{\textrm QI}$, 
where the transmission probability in one valley channel is strongly suppressed 
and the transport is therefore dominated by the other channel. 
This corresponds to a \emph{nearly perfect} valley filtering.

\subsection{Spin- and valley-split scenario} 
Lastly, we consider the case 
in which both the spin- and the chiral-$SU(2)$ symmetries are broken. 
It is more intuitive to take the spin-split scenario above as a starting point 
and break the chiral symmetry with the $\epsilon$ field. 
As shown in Fig.~\ref{fig:Te_evo}(d), 
the QI antiresonance of each valley splits under the effect of $\epsilon$, 
further reducing the degeneracy of $\omega^{\textrm QI}_{\lambda}$ to $g_{\lambda}\!=\!1$.
It is obvious that a completely equivalent description is obtained 
by taking the valley-split scenario as a starting point 
and increasing $U$ above $U^{AF}$ to induce magnetic order. 
Let us note that there is a non-trivial feedback 
between charge- and spin- correlations, 
resulting in a (weak) dependence of the critical threshold for spin ordering  
on the chiral symmetry-breaking field, 
i.e., $U^{AF}\!=\!U^{AF}(\epsilon)$. 
In fact, the formation of a charge-density wave requires 
to locally drive the C atoms away from half-filling. 
This is detrimental to the formation of the AF state, 
and it results in a partial quench of the local magnetic moments.~\cite{valliNL18} 
A consequence of this interplay is that, 
if the system is on the verge of magnetic ordering, 
tuning $\epsilon$ could allow to drive the system 
through a crossover between phases with different charge and spin order, 
and ideally working as a switch between spin-filtering 
and valley-filtering effects. 

\begin{figure}[b!]
\centering
\includegraphics[width=1.0\linewidth, angle=0]{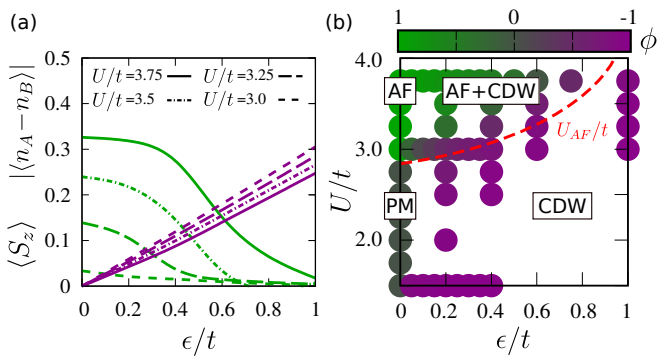}
\caption{(a) Evolution of the magnetic order parameters 
$\langle S_z \rangle$ (green lines) 
and the CDW order parameter $|\langle n_A\!-\!n_B \rangle|$ (violet lines)
with $\epsilon/t$ for different values of $U/t$. 
(b) Phase diagram identifying the PM and the AF (at $\epsilon\!=\!0$) 
as well as the CDW and the AF+CDW states. 
The quantity $\phi$ (see definition in the text)  
measures the weight of the different order parameters. 
The red dashed line separating the AF+CDW and the CDW states 
is to be intended as a guide to the eye 
indicating the crossover between the two states 
rather than a true estimate of $U_{AF}(\epsilon)$. }
\label{fig:op}
\end{figure}

Due to the finiteness of the system, it is not trivial 
to obtain a reliable estimate of $U_{AF}(\epsilon)$. 
There is however an alternative way to visualize this effect. 
In Fig.~\ref{fig:op}(a) we show the magnetic order parameter $\langle S_z \rangle$, 
which is suppressed by $\epsilon$ at any value of $U$. 
However, we note that it is difficult to reduce the order parameter 
below a certain numerical threshold 
since the observables for two spin components are evaluated independently. 
At the same time, the CDW order parameter $\langle n_A \!-\! n_B \rangle$ 
increases linearly with $\epsilon$. 
In Fig.~\ref{fig:op}(b) we show a phase diagram, 
characterized by the quantity 
\begin{equation}\label{eq:phi}
\phi \!=\! \frac{|\langle S_z \rangle| \!-\! |\langle n_A\!-\!n_B \rangle|}{|\langle S_z \rangle| \!+\! |\langle n_A\!-\!n_B \rangle|}, 
\end{equation}
which measures the relative weight of the AF and CDW order parameters. 
Hence, as $\phi(U,\epsilon) \rightarrow -1$,  
it highlights a crossover towards a pure CDW state (i.e., without AF order) 
and provides a reasonable estimate of $U_{AF}(\epsilon)/t$, 
as indicated by the dashed line (guide to the eye).

\begin{figure}[t!]
\centering
\includegraphics[width=1.0\linewidth, angle=0]{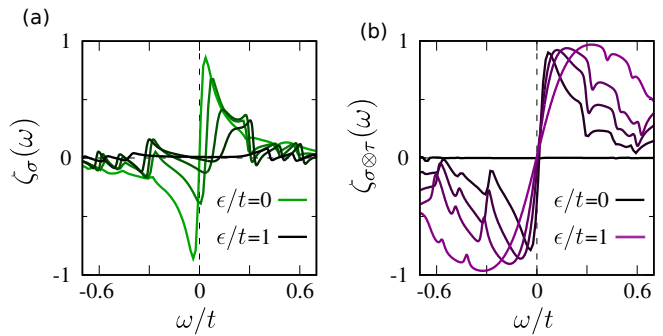}
\caption{Evolution with $\epsilon/t$ at $U/t\!=\!3$ 
of the spin-filtering efficiency $\zeta_{\sigma}(\omega)$ 
(for valley B, while the results for valley A can be obtained by symmetry) 
an the generalized valley-filtering efficiency $\zeta_{\sigma \otimes \tau}(\omega)$. 
Upon increasing the strength of the chiral symmetry-breaking field 
from $\epsilon/t\!=\!0$ to $\epsilon/t\!=\!1$ (in steps of $\Delta\epsilon\!=\!0.2$)   
$\zeta_{\sigma}(\omega)$ is suppressed (green to dark shades) 
while $\zeta_{\sigma \otimes \tau}(\omega)$ is enhanced (dark to violet shades) 
thus driving the system from a pure spin filter towards a pure valley filter. }
\label{fig:zevo}
\end{figure}

At the same time, the crossover from a spin filter to a valley filter 
can be observed directly by looking at the polarization of the transmission function 
across the $U_{AF}(\epsilon)$ line in parameter space. 
It is very intuitive to look at it as a function of $\epsilon/t$ 
at a fixed value of $U/t$, 
as shown in Figs.~\ref{fig:zevo}(a-b) for $U/t\!=\!3$. 
In Fig.~~\ref{fig:zevo}(a) we show the spin-filtering efficiency $\zeta_{\sigma}(\omega)$. 
At $\epsilon\!=\!0$ it displays two strong peaks at 
$\omega\!=\!\omega^{\mathrm QI}_{\sigma}$, 
which identify the destructive QI antiresonances 
causing the transport to be dominated by one spin channel. 
As $\epsilon/t$ increases both peaks are shifted to higher energy 
due to the inversion-symmetry breaking and are progressively suppressed 
as the AF order parameter $\langle S^z \rangle$ is suppressed, 
until the spin-filtering efficiency substantially vanishes. 
In order to quantify the valley filtering efficiency in the regime 
where both the spin-SU(2) and the valley-SU(2) symmetries are lifted, 
it is convenient to consider the quantity 
\begin{equation}\label{eq:zetasv}
 \zeta_{\sigma \otimes \tau} = 
 \frac{T_{A\uparrow} - T_{B\downarrow}}{T_{A\uparrow} + T_{B\downarrow}},
\end{equation}
which properly takes into account the fact that in the presence of AF order 
opposite valleys have also opposite spin polarization. 
In particular, $\zeta_{\sigma \otimes \tau}(\omega)\!\equiv\!0$ 
if $\epsilon/t=0$ even if $\langle S_z \rangle \!\neq\!0$, 
which is not true in the case of the quantity 
$\zeta_{\tau}(\omega)$ as defined in Eq.~(\ref{eq:zetav}). 
In Fig.~~\ref{fig:zevo}(b) we show that is strongly enhanced 
over an increasingly wider energy window as $\epsilon/t$ increases. 

Combining these two pieces of information demonstrates that, eventually, 
the chiral symmetry-breaking drives the system 
from a pure spin filter into a pure valley-filter.

%
%


\section{Origin of the QI antiresonances}
\label{sec:originQI}
To understand the mechanism leading to the $\omega^{\textrm QI}$ degeneracy lifting, 
we look explicitly at the structure of the Green's function. 

As already discussed, the general Landauer expression 
for the transmission function can be recast as in Eq.~(\ref{eq:trans_independent}), 
which establishes a direct link between $T(\omega)$ and $G_{\ell r}(\omega)$ 
for the generic $\ell\!\rightarrow\!r$ channel. 
In particular, in the WBL, all the frequency dependence of $T(\omega)$ 
comes from the Green's function. 
This means that a QI antiresonance (i.e., a zero of the transmission) 
necessarily implies a zero of the Green's function. 
At energies $|\omega|\!<\!\Delta$ (i.e., within the spectral gap) 
$\Im G_{\ell r}(\omega)\!\approx\!0$ for every pair $(\ell,r)$, 
where the exact relation holds at $T\!=\!0$. 
Therefore the zeroes of the Green's function
coincide with the zeros of $\Re G_{\ell r}(\omega)$. 
It can be shown (see Appendix~\ref{app:a1} for the derivation) 
that when neither the spin-$SU(2)$ nor the chiral symmetry is broken, 
the zero of the Green's function is pinned at the Fermi level ($\omega\!=\!0$)
by the particle-hole symmetry. 
Instead, any symmetry-breaking term shifts the zeros of $\Re G_{\ell r}(\omega)$, 
and hence the destructive QI antiresonance at finite frequency
(see Appendix~\ref{app:a2}).

\begin{figure}[t]
\includegraphics[width=0.47\textwidth, angle=0]{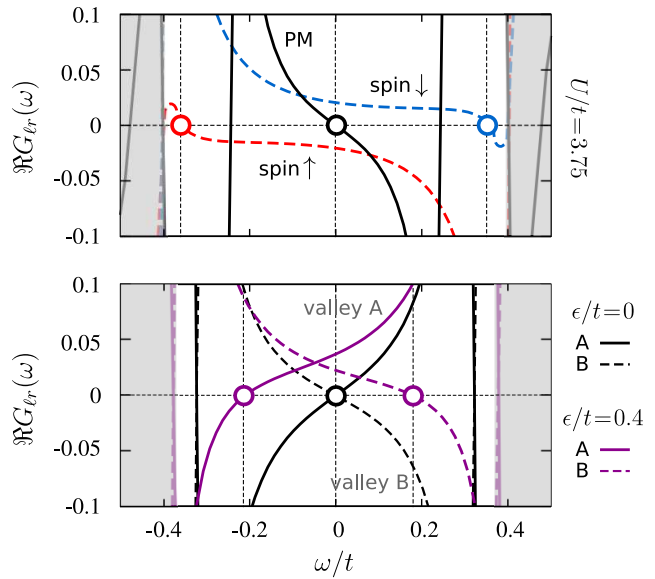}
\caption{(Color online) Zeros of the $\Re G_{\ell r}(\omega)$ 
for the sites in the middle of the edges 
of the meta configuration, without and with symmetry breaking. 
In the spin-split scenario for valley~A  (upper panel) and 
in the valley-split scenario for spin-$\uparrow$ (lower panel),
the two-fold degeneracy of the zero at $\omega_0\!=\!0$ is lifted, 
yielding two zeros at 
$|\omega_0^{\uparrow}|\!=\!|\omega_0^{\downarrow}|\!\approx\!\langle S^z\rangle\!\approx\!0.33$, 
and two zeros at 
$|\omega_0^{A}|\!=\!|\omega_0^{B}|\!\approx\!\langle n_B\!-\!n_A\rangle\!\approx\!0.19$, 
respectively. 
The grey shaded area indicates the energy window lying \emph{outside} 
the broken-symmetry gaps. 
Parameters: $U/t\!=\!3.75$, $\epsilon/t\!=\!0$, 
$\Gamma/t\!=\!0.02$ and $T\!=\!0.005t$ (upper panel); 
$U/t\!=\!1.5$, $\epsilon/t\!=\!0$ and $\epsilon/t\!=\!0.4$, 
$\Gamma/t\!=\!0.02$ and $T\!=\!0.005t$ (lower panel). }
\label{fig:originQI}
\end{figure}

In order to demonstrate this effect, 
in Fig.~\ref{fig:originQI} we explicitly show the low-energy structure 
of $\Re G_{\ell r}(\omega)$, for a given transmission channel, 
in which $\ell$ and $r$ are the sites 
in the middle of the $L$ and $R$ edges in the meta configuration (of sublattice B). 
The case of sublattice A can be obtained from this one by symmetry. 
In the upper panel of Fig.~\ref{fig:originQI} we show 
the effect of the spin-$SU(2)$ symmetry breaking. 
In the PM state, the zero of $\Re G_{\ell r}(\omega)$ 
is found at $\omega\!=\!0$ for both the spin-$\uparrow$ and spin-$\downarrow$ channels, 
while in the AF state we observe an opposite shift of the zeros to finite frequency 
which correlates with the behavior 
of the destructive QI antiresonance found at $\omega_{\sigma}^{\textrm QI}$, 
as shown in Fig.~\ref{fig:Te_evo}(c) (for sublattice B). 
In the lower panel of Fig.~\ref{fig:originQI} we demonstrate 
the analogous effect in for the chiral symmetry breaking. 
Contrary to the previous case, at $\epsilon=0$, 
$\Re G_{\ell r}(\omega)$ is not identical for the two valleys, 
but both display a zero at $\omega=0$. 
At finite field $\epsilon$, the zeros split symmetrically 
with respect to the Fermi level, yielding different 
$\omega_{\tau}^{\textrm QI}$ for the two valleys.

While an analytic expression for $\omega_{\lambda}^{\textrm QI}$ 
in the generic case cannot be easily obtained, 
in Appendix~\ref{app:a} we provide an argument 
that explains the relation between $\omega_{\lambda}^{\textrm QI}$ 
and the order parameters $\langle S^z \rangle$ and $\langle n_A-n_B \rangle$, 
which is shown in Figs.~\ref{fig:Te_spin} and \ref{fig:Te_valley} 
for the two symmetry-broken states, respectively. 
In particular, we stress that the shift of the zeros of the Green’s function is controlled by the low-frequency behavior of $\Re \Sigma_{ii}(\omega)$. 
This can be significantly different from the large-frequency one,  
which in turn can be seen as a sort of mean-field value. 
The difference between the low-frequency and high-frequency values 
of the self-energy can be taken as an estimate 
of the dynamical correlations beyond mean-field, 
and it has been shown to influence qualitatively the physics 
in models with a non-trivial band topology.\cite{amaricciPRL114,amaricciPRB98}




\section{Conclusions \& Outlook}
\label{sec:outlook}
We investigated the interplay between destructive QI and symmetry-breaking phenomena 
involving the spin and valley degrees of freedom in graphene nanoflakes. 
Specifically, by establishing a relation between the analytic structure 
of the real-space Green's function and the symmetries of the Hamiltonian, 
we provide a clear understanding of the origin of the QI antiresonances 
and of their effects on ballistic transport. 
Interestingly, our analysis works both in the symmetric 
and in the symmetry-broken cases, and we show that 
breaking a symmetry shifts the position of the antiresonance, 
without spoiling the destructive QI effects. 
This demonstrates the generality and the robustness 
of the phenomenon within a generic theoretical framework 
and also in the presence of electron-electron interactions. 
Ultimately, it allows us to predict the occurrence of QI antiresonances 
in complex nanostructures interacting with the environment. 

In our original proposal in Ref.~\onlinecite{valliNL18}, 
we showed that destructive QI 
can be used to generate nearly completely spin-polarized currents 
in the absence of magnetic fields or spin-orbit coupling, 
simply exploiting the spontaneous breaking 
of the $SU(2)$ spin-rotational symmetry induced by electronic correlation 
in the presence of the edges. 
The present work extends the scope of our previous study 
to multi-component systems where other degrees of freedom 
(e.g., valley, orbital, layer) 
can be manipulated via an external handle lifting the symmetry. 
In the specific case considered here, the mechanisms involved 
are the onset of a magnetically ordered state (associated with the electron spin) 
and the breaking of the inversion symmetry due to the interaction 
with a specific substrate (associated with the valley). 
We show that a tuning the coupling between the nanoflake and the substrate 
can turn a spin filter into a valley filter. 

The approach developed in the present work 
follows a general scheme according to which 
it is possible to manipulate the transport properties 
of a quantum junctions exploiting destructive QI, 
provided we have identified a symmetry 
and the corresponding symmetry-breaking control parameter. 
Other mechanisms suitable to this purpose are: 
(i) the switchable magnetic bistability of metal-organic complexes,~\cite{lefterMC2}
(ii) the Jahn-Teller distortions in charged fullerenes,~\cite{liuPRB97} and 
(iii) the formation of a moir\'{e} pattern 
in twisted bilayer graphene junctions.~\cite{rickhausNL18}
 
In this respect, we believe that our work 
can drive the community towards a promising  
and -so far- only sporadically explored direction.

\begin{acknowledgments}
We thank R.~Stadler and C.~Lambert for valuable discussions.  
We acknowledges financial support from MIUR PRIN 2015 (Prot. 2015C5SEJJ001) and SISSA/CNR project "Superconductivity, Ferroelectricity and Magnetism in bad metals" (Prot. 232/2015) 
and from the H2020 Framework Programme, 
under ERC Advanced Grant No. 692670 ''FIRSTORM". 
V.B. acknowledges support from Regione Lazio (L.R. 13/08) through project SIMAP. 
A.V. acknowledges financial support from the Austrian Science Fund (FWF) 
through the Erwin Schr\"odinger fellowship J3890-N36.
\end{acknowledgments}

\appendix

\section{Impact of symmetries on the real-space Green's function and the transmission}
\label{app:a}

\subsection{Symmetric state}
\label{app:a1}

At half-filling, the Hamiltonian of the flake ${\cal H}_F$ 
given in Eq.~(\ref{eq:HHM}) is symmetric under the following 
particle-hole transformation
\begin{equation}
\label{eq:phtrans}
\begin{aligned}
p^\dag_{Ai\sigma}&\rightarrow p_{Ai\sigma}& \ \ \ \ \
p_{Ai\sigma}&\rightarrow p^\dag_{Ai\sigma} \\
p^\dag_{Bi\sigma}&\rightarrow \ -p_{Bi\sigma}& \ \ \ \ \
p_{Bi\sigma}&\rightarrow \ -p^\dag_{Bi\sigma}.
\end{aligned}
\end{equation}

The pinning of the destructive QI antiresonance at the Fermi level 
can be demonstrated by considering the definition of the retarded Green's function
\begin{equation}\label{eq:Gr}
 G^r_{ij\sigma}(\o)=-i\int \theta(t) \lf\la  \lf\{p^{\phantom{\dagger}}_{i\sigma}(t), p^{\dagger}_{j\sigma}(0)\rg \}\rg\ra e^{-i\omega t} {\rm d}t ,
\end{equation}
where $\theta(t)$ is the Heaviside function, 
$\{\cdot,\cdot\}$ is the anticommutator for the fermionic operators, 
and the average is taken over the ground state at $T=0$, 
while it is replaced by the usual thermal average at $T\neq 0$. 
The invariance of the above expression under particle-hole transformation implies
\begin{equation}\label{eq:A}
G^r_{ij\sigma}(\o)=-i (-1)^{i+j} \int \theta(t) \lf\la  \lf\{p^{\dagger}_{i\sigma}(t), p^{\phantom{\dagger}}_{j\sigma}(0)\rg \}\rg\ra e^{-i\omega t} {\rm d}t 
\end{equation}
where the prefactor $(-1)^{i+j}$ equals $\pm 1$ depending on whether 
$i$ and $j$ belong to the same or to different sublattices. 
On the other hand, taking the complex conjugate of Eq.~(\ref{eq:Gr}) one obtains 
\begin{equation}\label{eq:Gstar}
\left[G^r_{ij\sigma}(\o)\right]^*=i\int \theta(t) \lf\la  \lf\{p^{\dagger}_{i\sigma}(t), p^{\phantom{\dagger}}_{j\sigma}(0)\rg \}\rg\ra e^{i\omega t} {\rm d}t .
\end{equation}
A comparison of equations Eq.~(\ref{eq:A}) and Eq.~(\ref{eq:Gstar}) 
demonstrates that in the presence of particle-hole symmetry 
the following relation holds
\begin{equation}\label{eq:phG}
\left[G^r_{ij\sigma}(\omega)\right]^*=(-1)^{i+j+1}G^r_{ij\sigma}(-\omega).
\end{equation}
This implies that, in the meta configuration, $\Re G_{ij\sigma}(0)$ 
is vanishing due to the particle-hole symmetry. 
Since $\Im G_{ij\sigma}(0)$ is suppressed by the presence of the spectral gap, 
the transmission from Eq.~(\ref{eq:trans_independent}) becomes 
\begin{equation}
 T^{\textrm{meta}}_{\sigma}(0) = 
 \sum_{\ell r} \Gamma_{\ell \ell} \Gamma_{rr} |\Im G^r_{\ell r\sigma}(0)|^2 
 \approx 0 ,
\end{equation}
where $\ell$ and $r$ span the proper subsets for the meta configuration. 
This demonstrates the pinning of the destructive QI at the Fermi level 
in the particle-hole symmetric case.
Moreover, this implies that a destructive QI is expected 
for any transmission channel $\ell\!\rightarrow\!r$ 
connecting sites from the same sublattice,\cite{pedersenPRB90,valliNL18} 
which allows us to \emph{predict} the occurrence of antiresonances 
in complex graphene nanostructures.

It is interesting to notice that the situation is drastically different 
in the other possible transport configurations. 
In both the ortho and para configurations, the sites 
$i$ and $j$ belong to different sublattices. 
Hence, Eq.~(\ref{eq:phG}) implies that $\Im G_{ij\sigma}(0)$ is vanishing, 
while $\Re G_{\ij\sigma}(0)$ is not. As a consequence, 
both $T^{\textrm{ortho}}_{\sigma}(0)$ and $T^{\textrm{para}}_{\sigma}(0)$ 
do not display any destructive QI antiresonance at the Fermi level.

Let us note that, in principle, an asymmetric coupling 
between the leads and the flake explicitly breaks the particle-hole symmetry.
However, in the WBL approximation, the leads introduce an additional lifetime 
$\Gamma_{\ell\ell}$ and $\Gamma_{rr}$, 
but they do not induce any energy shift to the poles, 
so that the excitation spectrum remains particle-hole symmetric. 
Furthermore, even in the case when the particle-hole is broken, 
the QI antiresonance would still exist 
at a finite frequency $\omega_{\lambda}^{\textrm QI}$.

\subsection{Symmetry-broken state}
\label{app:a2} 

In the presence of AF short-range order with a N\'{e}el pattern, 
the particle-hole transformation Eq.~(\ref{eq:phtrans}) 
has to be modified as follows, to leave the ground-state invariant
\begin{equation}
\label{eq:phtrans_af1}
\begin{aligned}
p^\dag_{Ai\sigma}&\rightarrow p_{Ai\bar{\sigma}}& \ \ \ \ \
p_{Ai\sigma}&\rightarrow p^\dag_{Ai\bar{\sigma}} \\
p^\dag_{Bi\sigma}&\rightarrow \ -p_{Bi\bar{\sigma}}& \ \ \ \ \
p_{Bi\sigma}&\rightarrow \ -p^\dag_{Bi\bar{\sigma}} 
\end{aligned}
\end{equation}
where $\bar{\sigma}=-\sigma$. 
In this case, Eq.~(\ref{eq:phG}) becomes 
\begin{equation}\label{eq:phG_af}
\left[G^r_{ij\sigma}(\omega)\right]^*=(-1)^{i+j+1}G^r_{ij\bar{\sigma}}(-\omega),
\end{equation}
which yields the following relation for the spin-dependent conductance 
\begin{equation}\label{eq:phT_af}
 T_{\sigma}(\omega)=T_{\bar\sigma}(-\omega)
\end{equation}
in all transport configurations (i.e., including also contact configurations 
of the nanoflake that do not exhibit destructive QI, 
such as the analogs of \emph{ortho} and \emph{para} configurations of benzene). 

Since the AF order and the graphene sublattices share the same real-space pattern, 
we can equivalently define the particle-hole transformation as 
\begin{equation}
\label{eq:phtrans_af2}
\begin{aligned}
p^\dag_{Ai\sigma}&\rightarrow p_{Bi\sigma}& \ \ \ \ \
p_{Ai\sigma}&\rightarrow p^\dag_{Bi\sigma} \\
p^\dag_{Bi\sigma}&\rightarrow \ -p_{Ai\sigma}& \ \ \ \ \
p_{Bi\sigma}&\rightarrow \ -p^\dag_{Ai\sigma} 
\end{aligned}
\end{equation}
where, with respect to Eq.~(\ref{eq:phtrans_af1}) we only exchanged 
the role of spin and sublattice indices. 

Let us now analyze the consequences of the invariance of the Green's function 
under the particle-hole transformation in Eq.~(\ref{eq:phtrans_af2}). 
When the Green's function connects sites belonging to different sublattices, 
as in the ortho and para configurations, 
the invariance under Eq.~(\ref{eq:phtrans_af2}) implies
\begin{equation}
 G^r_{ij}(\omega)=i\int \theta(t) \lf\la \lf\{p^{\dagger}_{j\sigma}(t), p^{\phantom{\dagger}}_{i\sigma}(0)\rg \} \rg\ra e^{-i\omega t} {\rm d}t
\end{equation}
that compared with Eq.~(\ref{eq:Gstar}) yields
\begin{equation}\label{eq:GABs}
\left[G^{r,AB}_{ij\sigma}(\omega)\right]^*=-G^{r,AB}_{ji\sigma}(-\omega).
\end{equation} 
where the superscript $AB$ indicate that $i$ and $j$ belong to different sublattices.
Eq.~(\ref{eq:GABs}) in turn implies for the total transmission in the ortho and para configurations
$T_{\sigma}(\omega)=T_{\sigma}(-\omega)$, 
and along with Eq.~(\ref{eq:phT_af}) 
eventually prevents the spin-filtering effect, yielding 
\begin{equation}
T^{\rm ortho}_{\sigma}(\omega)=T^{\rm ortho}_{\bar\sigma}(\omega), \ \ \ T^{\rm para}_{\sigma}(\omega)=T^{\rm para}_{\bar\sigma}(\omega).  
\end{equation}
On the contrary, in the meta configuration, similar reasoning shows 
that invariance under Eq.~(\ref{eq:phtrans_af2}) implies
\begin{equation}\label{eq:GAAs}
\left[G^{r,AA}_{ij\sigma}(\omega)\right]^*=G^{r,BB}_{ji\sigma}(-\omega)
\end{equation}
where the $AA$ and $BB$ superscripts indicate the two possible meta configurations, 
i.e., where only sites of sublattice A or only sites of sublattice B 
are connected to the leads.

Finally, Eq.~(\ref{eq:GAAs}) implies the following relation for the transmission
\begin{equation}\label{eq:Tss}
T^{\rm meta}_{\sigma AA}(\omega)=T^{\rm meta}_{\sigma BB}(-\omega), 
\end{equation}
that along with Eq.~(\ref{eq:phT_af}) yields
\begin{equation}\label{eq:Tssbar}
T^{\rm meta}_{\sigma AA}(\omega)=T^{\rm meta}_{\bar\sigma BB}(\omega) .
\end{equation}
Hence, provided that $\omega_{\sigma}^{\textrm{QI}}\!\neq\!0$, the above relations 
imply the spin- and sublattice- structure observed in the numerical simulations. 
Considering that in this case the AF and CDW order share the same real-space pattern, 
Eq.~(\ref{eq:Tss}) also demonstrates the properties of the chiral symmetry breaking case.

The last step of the analysis consists in the identification of the mechanism 
that shifts the QI antiresonance. 
Within DMFT, the spin-$SU(2)$ symmetry breaking is a spontaneous phenomenon. 
It is induced by the sort-range AF correlations due to the local repulsion $U$, 
resulting in a dynamical spin-dependent self-energy $\Sigma_{\sigma}(\omega)$. 
The static contribution of the self-energy 
$\Re \Sigma_{\sigma}(0) \propto \sigma \langle S^z \rangle$, 
acts as an effective chemical potential, 
with \emph{opposite} sign for the two spin polarizations, 
and it shifts the zeros of the Green's function to $\omega_{\sigma}^{\textrm{QI}}$. 
This effect is ultimately at the origin of the behavior 
observed in Fig.~\ref{fig:originQI} (upper panel) for the spin-split case. 
The chiral symmetry is instead explicitly broken by the field, 
so that the effective correction to the zero of the Green's function 
is given by $\epsilon \tau + \Re\Sigma_{\tau}(0) \propto \langle n_A-n_B \rangle$. 
Both terms have \emph{opposite} sign for the two valleys, 
and they induce the symmetric shift of the zeros to $\omega_{\tau}^{\textrm{QI}}$, 
as observed in Fig.~\ref{fig:originQI} (lower panel) for the valley-split case. 
Finally, in the spin- and valley-split case, the combination of the above  
self-energy corrections in the different channels 
can result in the complete lifting of the four-fold degeneracy of the QI antiresonance.

In general, the exact value of $\omega_{\lambda}^{\textrm{QI}}$ 
in a given transmission channel depends on the details 
of the real-space magnetization and charge redistribution pattern. 
Moreover, the transmission through the junction 
is in general given by the sum over the contributions of different channels, 
as shown in Eq.~(\ref{eq:trans_independent}). 
Hence, one might expect a distribution of antiresonances, 
one for each channel, 
which result in a broadening of the minima of the transmission  
with respect to the one pinned at the Fermi level 
and controlled by the particle-hole symmetry alone. 
Indeed, this effect is clearly observed in the numerical results.

\bibliographystyle{apsrev}
\input{arXiv2.bbl}

\end{document}

%% file: arXiv2.bbl
%